\documentclass{aa}

\usepackage{txfonts}
\usepackage{graphicx}
  \graphicspath{{figures/}}
\usepackage{pstricks}
\usepackage{multirow}
\usepackage{ulem}

\usepackage[authoryear,square]{natbib}

\usepackage[ps2pdf,linkbordercolor={1 1 1}, pdfborder={0 0 0}, colorlinks=true, linkcolor=black, citecolor=blue]{hyperref}

\begin{document}

%%%--- Informations on authors and institutes ---%%%
\title{Near-Infrared Mapping and Physical Properties of the Dwarf-Planet Ceres}

\author{Beno\^{i}t Carry\inst{1,2}
  \and Christophe Dumas\inst{1,3}\thanks{Affiliation at the time the observations were obtained}
  \and Marcello Fulchignoni\inst{2}
  \and William J. Merline\inst{4}
  \and J\'er\^{o}me Berthier\inst{5}
  \and Daniel Hestroffer\inst{5}
  \and Thierry Fusco\inst{6}
  \and Peter Tamblyn\inst{4}
}

\offprints{B. Carry, \email{bcarry@eso.org}}

\institute{
  ESO, Alonso de C\'{o}rdova 3107, Vitacura, Santiago de Chile, Chile
  \and LESIA, Observatoire de Paris-Meudon, 5 place Jules Janssens, 92190 Meudon Cedex, France
  \and NASA/JPL, MS 183-501, 4800 Oak Grove Drive, Pasadena, CA 91109-8099, USA
  \and SwRI, 1050 Walnut St. \# 300, Boulder, CO  80302, USA
  \and IMCCE, Observatoire de Paris, CNRS, 77 av. Denfert Rochereau, 75014 Paris, France
  \and ONERA, BP 72, 923222 Ch\^{a}tillon Cedex, France
  }

\date{Received 27 June 2007 / Accepted ...}

%%%--- Beginning of content ---%%%
\abstract
%---Context - OPTIONAL
{}
%---Aims
{We study the physical characteristics (shape, dimensions, spin  
axis direction, albedo maps, mineralogy) of
the dwarf-planet
Ceres based on high-angular resolution near-infrared  
observations.}
%---Methods
{We analyze adaptive optics J/H/K imaging  
observations of Ceres performed at Keck II Observatory in
September 2002 with an equivalent spatial resolution of
$\sim$50 km. The spectral behavior of the main geological features
present on Ceres is compared with laboratory samples.}
%---Results
{Ceres' shape can be described by an oblate spheroid ($a = b =
479.7 \pm 2.3$ km, $c = 444.4 \pm 2.1$ km) with EQJ2000.0 spin
vector coordinates $\alpha_0 = 288\degr \pm 5\degr$ and $\delta_0 =
+66\degr \pm 5\degr$. Ceres sidereal period is measured to be
$9.074\,10_{-0.000\,14}^{+0.000\,10}$ h.
We image surface features with diameters in
the 50-180 km range and
 an albedo contrast of $\sim$6\% with respect
 to the average Ceres albedo. 
The spectral behavior of the brightest regions on Ceres is
consistent with phyllosilicates and
carbonate compounds. Darker isolated regions could be related to the
presence of frost.}
%---Conclusion - OPTIONAL
{}


\maketitle

%%%---Various sections of the article ---%%%
  \section{Introduction}
%
%\footnotetext{$\star$ at time observations were obtained}
%
  \indent Ceres is by far the largest body among the population of
  main-belt asteroids. Curiously, although it was discovered more than
  200 years ago \citep{1802-Piazzi}, significant progress to
  understand its surface and interior properties has only been made
  over the last two decades. In particular we have seen a renewal of
  interest towards Ceres triggered by (i) the availability of
  sensitive spectro-imaging instrumentation on medium-to-large size
  telescope, enabling a detailed study of Ceres surface and physical
  properties \citep[e.g.][and others]{1996-Icarus-124-Mitchell,
  1998-Icarus-132-Drummond, 2000-AA-358-Dotto}, and (ii) more recently
  the selection of the NASA DAWN Discovery mission, which will visit
  Vesta and Ceres in 2011 and 2015 respectively
  \citep{2003-PSS-52-Russell}. But the main scientific interest drawn
  by Ceres is that it provides an excellent laboratory to understand
  how planetoids accreted early in the history of our solar system,
  and the role of volatiles in planetary formation and
  evolution. Unlike Vesta, which is dry and shows evidence of
  the melting phase and planetary differentiation
  \citep{2002-AstIII-Keil} seen in larger terrestrial planets like
  Earth, the ``dwarf planet'' Ceres (as it should now be named
  following the IAU guidelines) shows strong signs of water alteration
  on its surface \citep{1990-Icarus-88-Jones}. In this respect, Ceres
  displays stronger similarities with the icy outer satellites of
  Jupiter than with the dry asteroids that populate the inner region
  of the Main Belt. A possible scenario is that Ceres formed in a
  ``wet''   environment, from the accretion of both rocky
  planetesimals originally present at this heliocentric distance in
  the early planetary nebula, and icy planetesimals that migrated
  inward from the outer regions and whose ices had been preserved
  \citep{2005-MNRAS-358-Mousis}. As a consequence of this ``wet''
  history, the study of Ceres is of paramount importance to
  understand the process of planetary accretion and formation of the
  low-albedo primitive asteroids that populate the outer part of the Main
  Belt.\\
  \indent With a typical angular diameter of 0.6\arcsec~at opposition,
  Ceres can be spatially resolved from the ground using
  adaptive optics instruments available on medium-to-large
  telescopes. We thus carried out a program of multi-band imagery of
  Ceres from Keck Observatory at high-angular resolution, with
  the goals of precisely deriving its shape, dimensions,
  direction of spin axis, and distributions of albedo and
  color across its surface.
%
%  We thus carried out a high-contrast imaging program to
%  observe Ceres from Keck Observatory at high-angular resolution, and
%  precisely derive its shape, direction of spin axis, and distribution
%  of albedo features across its surface. The results of this
%  investigation are presented in details in this paper.
%

  \section{Observations and data reduction}
  \subsection{Observations}
%
%    Here we present results based on Ceres observations
%    performed at the Keck II Observatory on 22$^{\textrm{\tiny{nd}}}$ and
%    28$^{\textrm{\tiny{th}}}$ September 2002, one week before the opposition of
%    Ceres.
    \indent Our Ceres observations were made with the Keck II
        telescope on Mauna Kea, Hawaii, on 2002 September
        22 and 28 UT, one week before the opposition of Ceres.
    These observations were obtained under optimal atmospheric
    conditions with a sub-arcsec seeing at an airmass lower than 1.65
    (with half of the data taken with an airmass lower than
    1.2). Whereas the 2002 opposition occured near Ceres aphelia
    (Ceres was at a geocentric and heliocentric distance of 1.98 AU
    and 2.94 AU respectively), its apparent angular diameter was
    $\sim$0.66\arcsec. The phase angle was of 7\degr~and 5.5\degr~for
    the two nights, leading to an illuminated fraction of the surface
    of 99.6\% and 99.7\% respectively.\\
%
%    \indent Ceres was imaged using the second generation near infrared
%    1024x1024 InSb Aladdin-3 NIRC2 camera and the adaptive optics (AO)
    \indent Ceres was imaged using NIRC2,
    the second-generation near-infrared camera (1024x1024 InSb Aladdin-3) and
    the adaptive optics (AO)
    system installed at the Nasmyth focus of the Keck II telescope
    \citep{2004-AppOpt-43-vanDam}.
%
%    We obtained $9.942 \pm 0.05$
%    milliarcsec per pixel  images of Ceres at three near-infrared
%    wavebands J-, H- and K-band, that is [1.166,1.330] $\mu$m,
%    [1.485,1.781] $\mu$m and [1.948,2.299] $\mu$m
%    \textit{\textcolor{gray}{respectively}} (\textbf{Table \ref{tab-resolution_element}}),
%    inter-spaced with Point Spread Function (PSF) reference stars at
%    similar airmass and through the same set of filters.
%   
    The images of Ceres were acquired at 3 near-infrared
    wavebands J [1.166--1.330 $\mu$m], H [1.485--1.781 $\mu$m],
    and K [1.948--2.299 $\mu$m], with an image scale of
    9.942 $\pm$ 0.050 milliarcsec per pixel.
    Within the Ceres observation sequence we interspersed
    observations of reference stars, at similar airmass and
    through the same set of filters, to evaluate the system
    Point Spread Function (PSF).
    This calibration was required to perform \textsl{a posteriori} image
    restoration as described in the next section. No offset to sky was
    done but the telescope position was dithered between each exposure
    in order to record simultaneous sky
    and object frames, while the target (science or
    calibration) was positioned at three different locations on the
    detector, separated by nearly 5\arcsec~from each other. \\
  \subsection{Data reduction\label{subsec-data-reduction}}
\begin{figure}
%%%--- Graphics from the EPS and Caption
  \resizebox{\hsize}{!}{\includegraphics{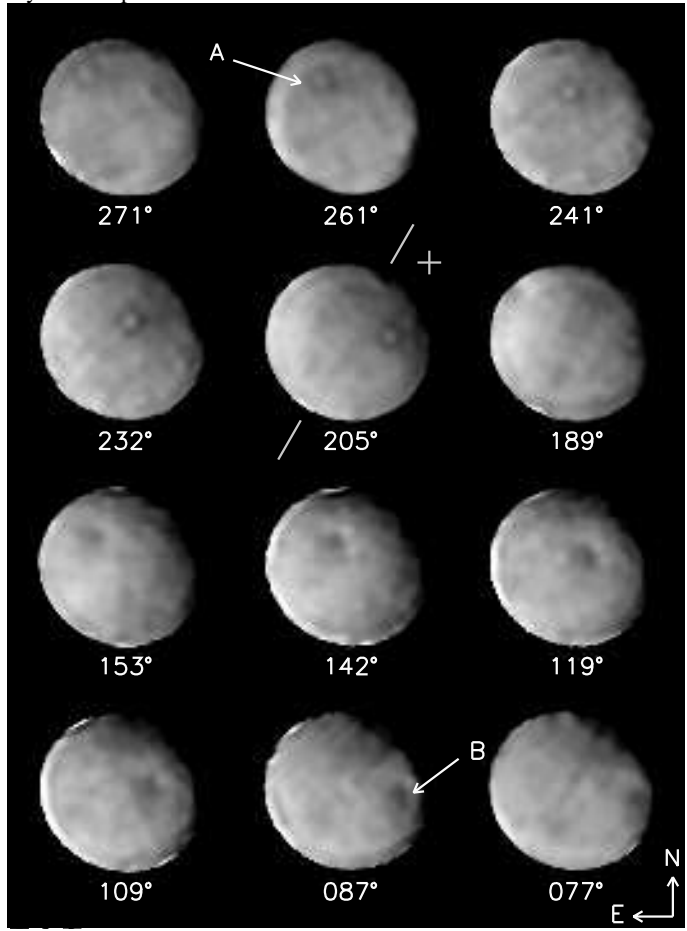}}

  \caption{Selected views of Ceres in K-band during $\sim$200\degr~of
  its rotation. The image is oriented with North up and East left. The
  values of the Sub-Earth Point longitude (SEP$_\lambda$) measured
  positively from 0 to 360 degrees in a right-hand system (following
  the IAU recommandation \citep{2005-CeMDA-91-Seidelmann}) are
  indicated below their corresponding images. Our meridian origin is
  chosen to be the same than \citet{2006-Icarus-182-Li}. The Ceres
  spin axis and positive pole are also indicated. The image stretch was chosen to enhance the surface features
  visible on Ceres at the detriment of the terminator. The two main
  surface features present on Ceres are indicated with the arrows A
  and B (see \ref{subsec-pole-coordinates}) and can be followed during
  part of their rotation. The brighter spots visible near the limb of
  Ceres in some of the images are artifacts from the deconvolution
  (see \ref{subsec-NIRmaps}).}
  \label{fig-views-of-ceres}

\end{figure}

    \indent We first reduced the data using the standard procedure for
    near-infrared images. A bad pixel mask was made by combining the
    hot and dead pixels found from the dark and flatfield frames. The
    bad pixels in our calibration and science images were then
    corrected by replacing their values with the median of the
    neighboring pixels (7x7 pixel box). Our sky frames were obtained
    from the median of each series of
    dithered science image, and then
    subtracted from the corresponding science images to remove
    the sky and instrumental background. By doing so, the dark current
    was also removed. Finally, each image was divided by a normalized
    flatfield to correct the pixel-to-pixel sensitivity
    differences of the detector. \\
%
%
%
%
%
%%%%%%%------ TABLE --- Begin --- Resolution Element ------%%%%%%
\begin{table}
  \centering
  \begin{tabular}{ccccccc}
    \hline\hline
     Filter & $\lambda_c$ & $\Delta \lambda$ & $\Theta$ & $\Theta$ Across & $\Theta$ Over \\
            & ($\mu$m)  & ($\mu$m)           &  (km)    & Diameter & Surface \\
    \hline
    J & 1.248 & 0.163 & 37.2 & $\sim$26 & $\sim$666 \\
    H & 1.633 & 0.296 & 47.9 & $\sim$20 & $\sim$400 \\
    K & 2.124 & 0.351 & 64.6 & $\sim$15 & $\sim$228 \\
    \hline
  \end{tabular}
  \caption{Central wavelength ($\lambda_c$) and bandpass width
    ($\Delta \lambda$) for each filter. The equivalent size (in
    km) of the theoretical resolution
    element ($\Theta$) on Ceres and the number of resolution elements
    across the diameter and over the apparent disk of Ceres are also reported. Ceres
    covers more than 3\,200 pixels (projected major- and minor-radius of
    $\sim$33 and $\sim$31 pixels respectively) on the NIRC2 detector.}
  \label{tab-resolution_element}

\end{table}
%%%%%%------ TABLE ---  End  --- Resolution Element ------%%%%%%
%
%
%
%
%
%
%
    \indent After these first basic reduction steps, we applied image
    deconvolution techniques to our set of Ceres data using the
    \textsc{Mistral} algorithm \citep{these-fusco, 2004-JOSAA-21-Mugnier}. The
    use of such an algorithm permits to restore the optimal spatial
    resolution of each image and is particularly well adapted to
    deconvolve objects with sharp edges, such as asteroids. Image
    restoration techniques are known to be
    constrained by the limitations to measure the precise instrumental +
    atmosphere responses at the exact same time the science observations
    are made. \textsc{Mistral} is a myopic deconvolution method, which estimates
    both the most probable object, and the PSF, from analysis of science
    and reference star images. The time needed for the algorithm to converge is
    largely dependent on the image size. Due to our large number of
    images to process, we decreased the deconvolution computation time by
    resizing all our images to a smaller (128x128 pixels, but still over 2x the diameter
      of Ceres) window centered
    on the object (Ceres or PSF). Comparison tests showed that no
    deconvolution artifacts were introduced by the use of smaller images. The
    deconvolved images of Ceres were then compared to identify
    the few frames (9 in total, that is less than 2.5\% of the full set
    of data) whose outputs were not satisfactory and discard them from our
    set of data. In the end we obtained 360 images of Ceres with a
    spatial resolution (\textbf{Table \ref{tab-resolution_element}})
    equivalent to the diffraction limit of a 10 m telescope (given by the
    angular sampling
    $\Theta = \lambda$/$D$, with $\lambda$
    the wavelength and $D$
    the telescope diameter). A subset of the restored images is presented
    in \textbf{Fig. \ref{fig-views-of-ceres}}.
%
%
%
%
%
%
%%%%%%------ FIGURE --- Begin --- Ceres during its rotation ------%%%%%%
%%%%%%------ FIGURE --- End --- Ceres during its rotation ------%%%%%%

  \section{Ceres global physical properties\label{sec-analysis1}}
  \subsection{Spin vector coordinates\label{subsec-pole-coordinates}}
%
%
%%%%%%------ FIGURE --- Begin --- Control Point Method ------%%%%%%
\begin{figure}
  \begin{center}
    \resizebox{0.6\hsize}{!}{\includegraphics{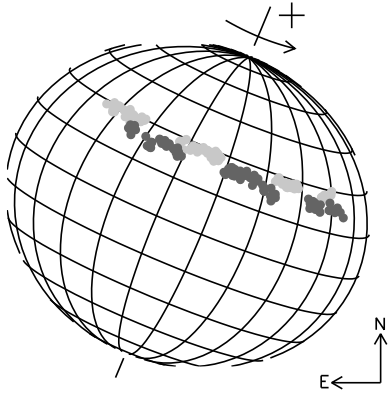}}\\

    \caption{Apparent motion of the two main features A (bright) and B
    (dark) over $\sim$90\degr~of Ceres rotation during the night of
    the 2002 September 22 UT (SEP$_\varphi \simeq
    +8$\degr, $p_n \simeq 338$\degr). We used the rotational track of
    these two features to determine the direction of the rotation axis
    (as described in text).}
   \label{fig-CPM}
  \end{center}
\end{figure}
%%%%%%------ FIGURE --- End --- Control Point Method ------%%%%%%
%
%
%
    \indent Measuring the Spin Vector Coordinates (SVC) of Ceres has
    always been a difficult task. The small amplitude of its
    lightcurve ($\sim$0.04 mag) prevented its determination using
    classical lightcurve inversion techniques. Until recently, the
    solutions that were reported for the pole coordinates of Ceres
    were widely dispersed (spanning a 90\degr~range in the plane of
    the sky during the 1995 opposition as reported by
    \citet{2002-AJ-123-Parker}, based on the
    compilation of 
    \citet{1983-Icarus-56-Johnson},
    \citet{1987-Icarus-72-Millis},
    \citet{1993-Icarus-105-SaintPe1} and 
    \citet{1998-Icarus-132-Drummond} pole solutions). This situation changed dramatically
    with the use of high-contrast direct imaging techniques capable of
    resolving the disk of Ceres and image fine details on its surface, such
    as AO observations (e.g \citep{2003-DPS-Dumas}, and this paper)
    and HST observations (e.g. \citep{2002-AJ-123-Parker,
    2005-Nature-437-Thomas}). The high-quality data obtained at Keck allow us to obtain an
    independent measurement of the SVC of Ceres, which is compared to
    the HST results \citep{2005-Nature-437-Thomas}. Precise knowledge
    of the SVC is mandatory to re-project the individual images into
    Ceres planetocentric referential and establish albedo maps of its surface (see \ref{subsec-NIRmaps}). \\
    \indent We performed a control-point method, which consists of
    following the apparent motion of a surface feature while the
    asteroid rotates. The path of a rotating feature on the surface of
    Ceres describes an ellipse if the asteroid is seen under a
    Sub-Earth Point latitude (SEP$_\varphi$) different from 0\degr, or
    $\pm$90\degr~(where it describes a line, or a circle respectively). We
    marked the positions of two main features A and B
    (\textbf{Fig. \ref{fig-views-of-ceres}}) while they rotated
    respectively over a 93.9\degr~and 87.7\degr~range around Ceres'
    spin axis (100 and 87 images respectively). All these measurements are
    shown together on a 3-D representation of Ceres in
    \textbf{Fig. \ref{fig-CPM}}. The shape of
    the projected track being determined by
    the direction of the rotational axis, we were able to measure
    the north pole angle ($p_n$) and the SEP$_\varphi$ at the time of
    the observations.\\
    \indent In order to compare our solution with other results, we
    followed the IAU recommendations and calculated the SVC in the
    J2000.0 equatorial frame: $\alpha_0 = 288\degr \pm 5$\degr~and
    $\delta_0 = +66\degr \pm 5$\degr. This result is in agreement with
    the latest result from HST \citep{2005-Nature-437-Thomas}
    ($\alpha_0 = 291\degr \pm 5$\degr~and $\delta_0 = +59\degr \pm
    5$\degr).\\
    \indent To directly derive the obliquity of a planetary body, it
    is also convenient to express the SVC in an ecliptic reference
    frame. Our measurements give $\lambda_0 = 7$\degr~and $\beta_0 =
    +83$\degr~(with 5\degr~uncertainty), leading to an obliquity for
    Ceres of about 4\degr. Such a small obliquity was expected from
    analysis of lightcurve data \citep{1983-Icarus-54-Tedesco} and
    thermal properties of Ceres regolith
    \citep{1990-Icarus-83-Spencer}. Indeed, a larger obliquity would
    imply stronger variations in the amplitude of Ceres' lightcurves,
    which have never been reported. As a result, Ceres can only be
    observed over a small range of Sub-Earth Point latitude
    (SEP$_\varphi$) [-11\degr, +10\degr] (computed until 2015),
    imposing on us to wait for the arrival of the
    DAWN spacecraft to get a direct view of its polar regions.
 \subsection{Rotational period\label{subsec-rotation}}
%
%%%%%%------ TABLE --- Begin --- Rotation ------%%%%%%
\begin{table}
  \centering
  \begin{tabular}{c c c}
    \hline\hline
    Epoch & Date         & SEP$_\lambda$ \\
          & (Julian Day) &  (\degr)      \\
    \hline
    t$_1$ & 2\,452\,539.894\,02   & 234.6 $\pm$ 2  \\
    t$_2$ & 2\,452\,545.915\,46   & 261.4 $\pm$ 2  \\
    t$_3$ & 2\,453\,002.241\,28   & \textcolor{white}{0}22.6 $\pm$ 5  \\
    \hline
  \end{tabular}
  \caption{Sub-Earth Point longitude (SEP$_\lambda$) for the three
  epochs (t$_1$ and t$_2$ from Keck, t$_3$ from HST). Using t$_1$
  as reference, Ceres sidereal phasing with $t_2$ and $t_3$ can be
  obtained by applying a
  +1.3\degr~and a -98.7\degr~correction taking into account the apparent
  geometry of the Earth  with respect to Ceres.}
  \label{tab-JD-SEPlong}
\end{table}
%%%%%%------ TABLE ---  End  --- Rotation ------%%%%%%
%
    \indent We establish a precise and independent
    measurement of the rotation period of Ceres using images from our two Keck epochs
    (2002 September 22 and 28 UT), plus
    one additional processed Hubble
    ACS/HRC image (\textsl{j8p502amq\_iof.fit}) from the HST program GO
    9748 retrieved from the Small Body Nodes archive
    \citep{DATA-HST-Ceres2003-04} for
    the epoch 2003 December 28 UT (\textbf{Table
    \ref{tab-JD-SEPlong}}) \citep{2005-Nature-437-Thomas, 2006-Icarus-182-Li,
    2006-AdSpR-38-Parker}. Two main
    albedo marks visible in all images were used to obtain precise information
    on Ceres rotational phase at these three epochs (see \textbf{Table
    \ref{tab-JD-SEPlong}}). In addition, the period between the Keck
    and HST observations was sufficiently large to accurately measure
    the error accumulation on Ceres' period over more than 1000 rotations.\\
%
%%%%%%------ TABLE --- Begin --- Rotation Results ------%%%%%%
\begin{table}
  \centering
  \begin{tabular}{c c c c c}
    \hline\hline
    Sidereal period & SEP$_\lambda$(t$_2$) & $\Delta$SEP$_\lambda$ & SEP$_\lambda$(t$_3$) & $\Delta$SEP$_\lambda$ \\
    (h) &  (\degr) & (\degr) &  (\degr) &  (\degr) \\
    \hline
    9.066\,588        &  255.1  &  ~6.3  &  17.9  &  ~4.7  \\
    9.066\,685        &  255.1  &  ~6.3  &  22.6  &  ~0.0  \\
    9.066\,780        &  255.1  &  ~6.3  &  27.2  &  -4.6  \\
    \hline
    9.074\,000        &  259.8  &  ~1.6  &  17.8  &  ~4.8  \\
    9.074\,090        &  259.8  &  ~1.6  &  22.1  &  ~0.5  \\
    \textbf{9.074\,100}&  \textbf{259.8}  &  ~\textbf{1.6}  &  \textbf{22.6}  &  ~\textbf{0.0} \\
    9.074\,110        &  259.8  &  ~1.6  &  23.1  &  ~0.5  \\
    9.074\,200        &  259.9  &  ~1.6  &  27.5  &  -4.9  \\
    \hline
    9.081\,526        &  264.4  &  -3.0  &  17.9  &  ~4.7  \\
    9.081\,526        &  264.5  &  -3.1  &  22.6  &  ~0.0  \\
    9.081\,526        &  264.6  &  -3.2  &  27.2  &  -4.6  \\
    \hline
  \end{tabular}
  \caption{Sidereal periods of Ceres which are in agreement with our
  observations, based on SEP$_\lambda(t_1) = 234.6$\degr. The predicted
  Sub-Earth Point longitude (SEP$_\lambda$) are computed at epochs
  t$_2$ and t$_3$ and are compared with the measurements (difference
   $\Delta$SEP$_\lambda$). We only kept sidereal periods whose predicted
  SEP$_\lambda$(t) were inside measurement uncertainties.}
  \label{tab-RateVSecart}

\end{table}
%%%%%%------ TABLE ---  End  --- Rotation Results ------%%%%%%
%
    \indent We then used the Eproc ephemeris generator
    \citep{1998-IMCCE-Berthier} to predict the longitude of the
    Sub-Earth Point and determine the value of the Ceres sidereal
    period that minimizes the difference $\Delta$SEP$_\lambda$ between the
    observed and the computed SEP$_\lambda$. We adopted the pole
    solution derived in this work as well as the priorly
    determined Ceres sidereal period of $P_s = 9.075 \pm 10^{-3}$ h
    \citep{1983-Icarus-54-Tedesco} as a best estimate.
    \textbf{Table
    \ref{tab-RateVSecart}} shows the sidereal periods that are in
    agreement with the Sub-Earth Point longitude (SEP$_\lambda$) at
    the epochs of the observations.\\
    \indent The value of Ceres sidereal period best matching our
    observations was found to be $P_s =
    9.074\,10_{-0.000\,14}^{+0.000\,10}$ h with a $\sim 10^{-4}$ h
    resolution. 
    The small $\sim$0.5 s uncertainty, which come mainly from the
    error on the mesured longitudes (see \textbf{Table \ref{tab-JD-SEPlong}}),
    makes possible to predict Ceres' Sub-Earth Point longitude,
    SEP$_\lambda$, with an error of only 40\degr~in rotational phase
    over the next decade. By comparison, the very recent study by
    \citet{2007-Icarus-188-Chamberlain} compiled 50 years of
    lightcurve measurements to derive a period of $P_s = 9.074\,170 \pm
    2 \times 10^{-6}$ h. \\
%
%
%
%
%
%%%%%%------ TABLE --- Begin --- Rotation ------%%%%%%
\begin{table}
  \centering
  \begin{tabular}{c c c c c}
    \hline\hline
    $\alpha_0$ & $\delta_0$ &  $W_0$   & $P_s$ &  Epoch\\
     (\degr)   & (\degr)    &  (\degr) & (h)   &  (JD)\\
    \hline
    $288 \pm 5$      & $+66 \pm 5$     & $-46 \pm 2$  & $9.0741 \pm 10^{-4}$ & 2\,452\,539.894\,02 \\
    \hline
  \end{tabular}
  \caption{Ceres sidereal period ($P_s$), pole direction ($\alpha_0$,
    $\delta_0$) and initial rotational phase angle ($W_0$) at epoch of
    reference in the J2000.0 equatorial frame.}
  \label{tab-PoleSolution}
\end{table}
%%%%%%------ TABLE ---  End  --- Rotation ------%%%%%%
%
%
%
%
%
%
%
%
%
%
%
%
%
  \subsection{Dimensions\label{subsec-dimensions}}
\begin{figure}
  \begin{center}
%%%--- Graphics from the EPS and Caption
    \resizebox{.49\hsize}{!}{\includegraphics{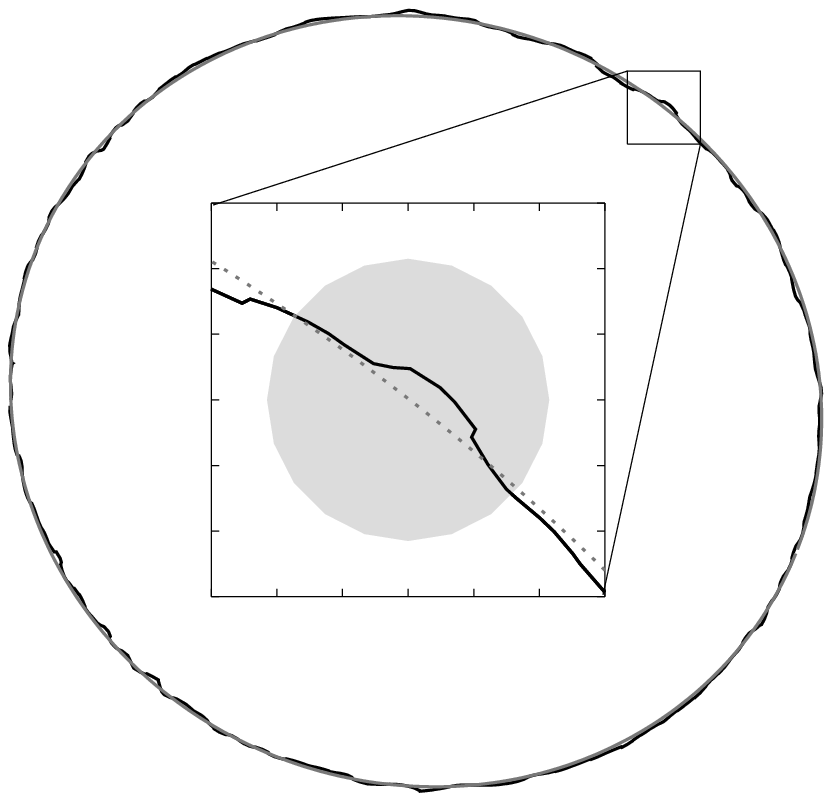}}
    \resizebox{.49\hsize}{!}{\includegraphics{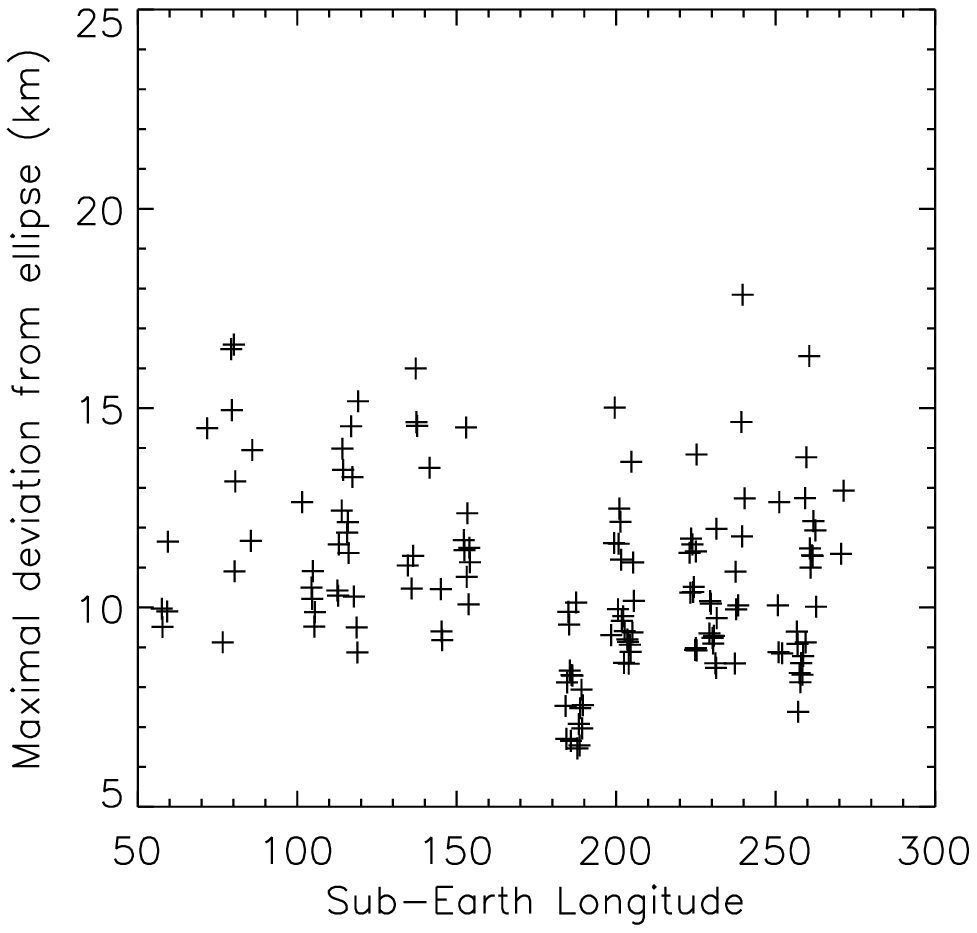}}
    \caption{Left figure: Extracted limb contour (black) from a J-band
      image of Ceres with its ellipsoidal fit overplotted for
      comparison (dotted gray). This example shows that deviations
      from the ellipsoidal model are much smaller than the resolution
      element (gray disk inside the box) obtained with Keck-AO. Right
      figure: Compilation of our deviation measurements as function of
      the Sub-Earth Point longitude (SEP$_\lambda$). No deviations
      larger than $\sim$18 km (half a resolution element at J-band)
      are detected.}
    \label{fig-edge}
  \end{center}
\end{figure}

    \indent Precise measurements of the shape of Ceres allow us to
    remotely investigate its internal structure and test whether it is
    a differentiated body \citep{2005-Nature-437-Thomas}. We thus
    performed a Laplacian of Gaussian (LoG) wavelet analysis to 192
    images of Ceres to extract its limb contours. Because it is very
    sensitive to variations of gradient, the LoG permits to precisely
    detect the inflection points in the flux distribution of our
    individual deconvolved images of Ceres. We found Ceres to be
    rotationally symmetric as first reported from a preliminary
    analysis of the Keck-AO data by \citet{2003-DPS-Dumas}, and later
    confirmed by \citet{2005-Nature-437-Thomas}. Analysis of our
    complete set of contours did not return any deviation from our
    ellipsoidal model larger than $\sim$18 km (see
    \textbf{Fig. \ref{fig-edge}}). From this model, and our knowledge
    of the direction of its spin axis, we were able to determine the
    minor- and major- projected radius of Ceres and correct them from
    their aspect and phase angles. We found that Ceres is well
    described by an oblate spheroid whose semi-axes are $a = b = 479.7
    \pm 2.3$ km and $c = 444.4 \pm 2.1$ km (1-$\sigma$ dispersion for
    the fitted axis). These values are different from HST's by
    $\sim$10 km (relative difference of two percent). In comparison,
    the agreement between our semi-major axis and the determination
    made from stellar occultation by \citet{1987-Icarus-72-Millis} is
    remarkable: $479.6 \pm 2.4$ km (occultation) \textsl{vs} $479.7
    \pm 2.3$ km (Keck), while our minor-axis value differ from theirs
    by $\sim$6-9 km (\textbf{Table \ref{tab-KECKvsHST}}). Whereas
    \citet{1987-Icarus-72-Millis} assumed a zero-obliquity at the time
    of the occultation, our SVC solution gives a SEP$_\varphi$ of
    +3.3\degr. The corrected polar radius from stellar occultation is
    thus $453.3 \pm 4.5$ km, which is still different from our
    measurement by nearly 10 km. It is important to note that during
    the occultation of 1987, the cords along Ceres orbit were aligned
    in an East-West direction, whereas the north pole position angle
    was $p_n \sim 339$\degr. This implies that the estimation of the
    small radius of Ceres was more loosely constrained than its
    semi-major axis. In addition, \textbf{Table \ref{tab-KECKvsHST}}
    provides a comparison between the pixel size of the HST and Keck
    images, and shows that the coarser sampling of the disk of Ceres
    as seen by HST could explain the differences in our respective
    determinations of its size. Our mean radius for Ceres is estimated
    to be $R=467.6 \pm 2.2$ km, which is also closer to the
    determination made from stellar occultation than HST's. \\
%
%%%%%%------ FIGURE --- Begin --- Edge Detection ------%%%%%%
%%%%%%------ FIGURE --- End --- Edge Detection ------%%%%%%
%
%
%%%%%%------ TABLE --- Begin --- Keck-HST-Occultation ------%%%%%%
\begin{table}
  \centering
  \begin{tabular}{ccccc}
    \hline\hline
    Quantity          & \multicolumn{2}{c}{Imagery} & \multicolumn{2}{c}{Occultation}         \\
      (km)            &     Keck          &         HST           &      Sol.1          &     Sol.2         \\ 
    \hline
    $a$               &  479.7 $\pm$ 2.3  &   487.3 $\pm$ 1.8     &   479.6 $\pm$ 2.4   &  481.6 $\pm$ 2.4  \\
    $c$               &  444.4 $\pm$ 2.1  &   454.7 $\pm$ 1.6     &   453.4 $\pm$ 4.5   &  450.1 $\pm$ 2.0  \\
%    $R=\sqrt[3]{aac}$ &  467.6 $\pm$ 2.2  &   476.2 $\pm$ 1.7     &   470.7 $\pm$ 3.1   &  470.8 $\pm$ 2.3  \\
    $R$ &  467.6 $\pm$ 2.2  &   476.2 $\pm$ 1.7     &   470.7 $\pm$ 3.1   &  470.8 $\pm$ 2.3  \\
    pixel             &  \textcolor{white}{0}14.3 $\pm$ 0.7  &   33.9 $\times$ 29.6  &  ... &... \\
    \hline
\end{tabular}
  \caption{Semi-major ($a$) and -minor ($c$) axes, and equivalent
    radius ($R$=$\sqrt[3]{aac}$) for Ceres derived in this paper (Keck),
    \citet{2005-Nature-437-Thomas} (HST), and
    \citet{1987-Icarus-72-Millis} (Occultation). The pixel size of the
    Keck and HST images are also given for comparison.}
  \label{tab-KECKvsHST}
\end{table}
%%%%%%------ TABLE ---  End  --- Keck-HST-Occultation ------%%%%%%
%
%
%
%
%
    \indent Adopting a mass for Ceres of $M = 9.43 \pm 0.07 \times
    10^{20}$ kg (average of most recent measurements
    \citep{1998-AA-334-Viateau, 2000-AA-360-Michalak,
      2007-EMP-100-Kovacevic}), we find a mean density $\rho=2\,206 \pm
    43$ kg.m$^{-3}$. This value is relatively high for a hydrated
    G-type asteroid like Ceres, but can be explained by a low porosity
    \citep[see][]{2002-AstIII-Britt}, and is similar to the density of
    the icy outer Jovian satellites Ganymede and Callisto. One can
    assume Ceres to be in hydrostatic equilibrium and inverse the
    relation between $a$, $c$ and $J_2$~\citep[given
      by][]{2005-JGR-110-McCord} as following
$$
      J_2 = \left[1 - \frac{c}{a} - \frac{2 \pi^{2} R^{3} }{P_s^{2} G M} \right]\left[ \frac{c}{2a} + \left( \frac{a}{c} \right)^{2} \right]^{-1}
$$
    \noindent and find $J_2=26.7 \times 10^{-3} \pm 1.9 \times
    10^{-5}$. By comparison, \citet{2005-Nature-437-Thomas} have found
    $J_2=21.7 \times 10^{-3} \pm 8.5 \times 10^{-5}$. If we refer to
    \cite{2005-JGR-110-McCord}, these two independent estimates of
    $J_2$ correspond to their internal models \#2 and \#3 for Ceres,
    that is to a differentiated Ceres with a silicate-rich region in
    its center. The rather large difference between the two $J_2$
    determinations is due to the fact that $J_2$ is highly dependant
    on the $a$/$c$ ratio which shows large uncertainties depending on the technique used for 
    its measurement. Lightcurve analysis for a quasi-spherical object is 
    generally little sensitive to this parameter; while measurements made 
    from direct imaging from the Earth distance suffer from lack of
    sufficient spatial resolution to estimate this 
    parameter with the required precision.
    Nevertheless, both studies converge towards a
    differentiated asteroid and gravity field measurement made by DAWN
    will provide a better understanding of its mass repartition and
    internal structure. \\
    \indent \citet{2007-Icarus--Conrad} have shown that a detailed
    study of an asteroid shape, and its departure from a perfect
    ellipsoid, can be carried out using images deconvolved with \textsc{Mistral}.
    Deviations of Ceres limb measurements
    from our shape model can thus directly be linked to
    topography. \textbf{Fig. \ref{fig-edge}} shows that no features
    deviating from our shape model by more than 15-18 km can been
    observed.
    The highest relief expected on Ceres is calculated to be $\sim$10-20
    km high \citep[see also][Fig. 1]{1973-Icarus-18-Johnson},
    thus confirming that no significant deviation from an ellipsoid
    can be detected given the size of our resolution
    element. Search for relief would require a resolution of
    at least about 5 km to provide unambiguous detection of a topography.

\section{Ceres surface composition\label{sec-analysis2}}
  \subsection{Background\label{sec-review-water}}
%
%%%---- H2O at Ceres heliocentric distance is possible
    \indent Ceres occupies a particular place in our solar system. It
    is physically located far enough from the Sun to have been
    preserved from strong heating during the T Tauri phase of the Sun
    \citep[][and references therein]{1988-Meteorites-Ghosh},
    and has possibly retained some of its primordial elements. Its low
    amplitude lightcurve ($\sim$0.04 mag), which cannot result from
    Ceres’ symmetrical shape \citep{2005-Nature-437-Thomas}, shows
    that its surface, while rather uniform, displays faint albedo
    features, unlike the igneous asteroid Vesta whose hemispheric
    albedo variations are among the strongest seen among main belt
    objects \citep{1997-Icarus-127-Gaffey}. Similarly, in contrast to Vesta, whose shape is highly
    irregular due to past collisions \citep{1997-Science-277-Thomas},
    Ceres displays a uniform spheroidal shape, deprived of strong
    surface and topographic features, and its density (see
    \ref{subsec-dimensions}) cannot simply be attributed to
    macroporosity of its internal material
    \citep[see][]{2002-AstIII-Britt}. All these characteristics point
    to the presence of volatile elements in the interior of
    Ceres. \citet{2005-MNRAS-358-Mousis} show that Ceres could have
    accreted from an assemblage of rocky and icy planetesimals, even
    at such short heliocentric distances. They calculated that icy
    planetesimals could have drifted from more distant regions of our
    solar system (up to 15 AU heliocentric) to the actual position of
    Ceres (2.7 AU) without losing entirely their volatiles. This idea
    supports the model of \citet{1989-Icarus-82-Fanale} in which the
    C/G-type objects accreted from anhydrous minerals, organics and
    water ice. The recent discovery of comets orbiting among the
    main-belt asteroids, at semi-major axes similar to Ceres'
    \citep{2006-Science-312-Hsieh} support the possibility that small
    bodies can preserve part of their ices within
    the snow line region, which is
    defined as the heliocentric distance for which the temperature equals
    the condensation temperature of water ($\sim$5 AU).\\
%
%%%----Observations in favor or versus H2O on Ceres
    \indent Other studies supporting a wet history for Ceres come from
    meteorites. Although no meteorites have been convincingly
    linked to Ceres 
    \citep{1981-GeCoA-45-Feierberg,1990-Icarus-88-Jones,1997-MPS-32-Sato}, the C-type asteroids (whom G-type is a subclass)
    display a low albedo, and are thought to be the source of the Carbonaceous 
    Chondrites (CC) meteorites. It has been shown that water ice could be stable
    inside CC meteorites over 4.5 Gyr \citep[][and references therein]{1989-Icarus-82-Fanale},
    thus supporting the possibility that the Main
    Belt has likely experienced a hydrated stage in its history 
    (at least its outermost part). Moreover, spectral studies of Ceres reveal 
    a strong 3.07 $\mu$m absorption band characteristic of hydrated minerals
    \citep{1990-Icarus-88-Jones,1997-MPS-32-Sato}. This particular band has been the center of interest of many studies: In the early
    80's, \citet{1981-GeCoA-45-Feierberg} and
    \citet{1981-Icarus-48-Lebofsky} associated it with the signature
    of water frost on Ceres surface and predicted the possible
    existence of a polar cap. A decade later,
    \citet{1992-Science-255-King} fitted the 3 $\mu$m absorption with
    saponite, which is an ammonium-bearing phyllosilicate whose presence in
    CV and CI meteorites have been suggested by
    \citet{1988-Meteorites-Zolensky}.
    Later, \citet{2005-AA-436-Vernazza} found
    that crystalline water ice mixed with ion-irradiated asphaltite
    could reproduce a better fit to this feature. Recently,
    \citet{2006-Icarus-185-Rivkin} reviewed this past work and found
    that an hydrated iron-bearing phyllosilicate identified as
    cronstedtite, plus a few percent of carbonates could also fit
    adequately the near-infrared spectrum of Ceres. This recent
    interpretation could be supported by the detection in the mid-infrared
    range of emission features attributed to carbonates
    \citep[from][]{1998-AJ-115-Cohen}. In addition, iron-bearing
    minerals have also been invoked by \citet{1981-GeCoA-45-Feierberg}
    and \citet{1989-Science-246-Vilas}
    to explain other spectral
    features present in the visible and near-infrared spectra of
    Ceres: 0.4 $\mu$m cutoff, 0.60 $\mu$m and 0.67 $\mu$m weak absorption bands
    and  1 $\mu$m shallow band \citep[see also][]{1992-Icarus-100-Vilas}.
    Given the radical different interpretations of the $3.07$
    $\mu$m band reported over the past decade, it is quite impossible
    to conclude on its exact nature. A persistent conclusion though,
    consists to support the presence of hydrated minerals, or residues
    from aqueous alteration, in the regolith material. \\
%
%%%---- Why not some ice in short time scale ?
    \indent As discussed above, surface water ice is not stable at
    distances smaller than 5 AU and is expected to sublimate if
    exposed directly to solar radiation
    \citep{1989-Icarus-82-Fanale}. Indeed, water ice migrating from the
    mantle region could possibly reach the surface but will escape on
    very short time-scale \citep{2003-AdSpR-31-Nazzario}. As a result,
    direct detection of water ice vaporization from Ceres surface
    might be possible from the surroundings of a fresh impact crater,
    or from cracks in the sub-surface layers. Water escaping from Ceres polar
    region has possibly been detected in the early 90's
    \citep{1992-Icarus-98-AHearn}, although this 2-$\sigma$ detection could never
    be unambiguously confirmed. Another supporting
    element comes from the relaxed shape of Ceres, which could be
    explained by the presence of large amount of ice in its interior
    \citep{2005-JGR-110-McCord}. The measurements of $J_2$ obtained
    from HST and Keck both clearly support a model of a differentiated
    Ceres with a volatile-rich mantle, rather than a homogeneous
    interior model (see \ref{subsec-dimensions}).\\
  \subsection{Near-infrared maps\label{subsec-NIRmaps}}
%
%%%%%%------ FIGURE --- Begin --- J-/H-/K-band Maps ------%%%%%%
\begin{figure*}[!ht]
  \centering
  \includegraphics[width=17cm]{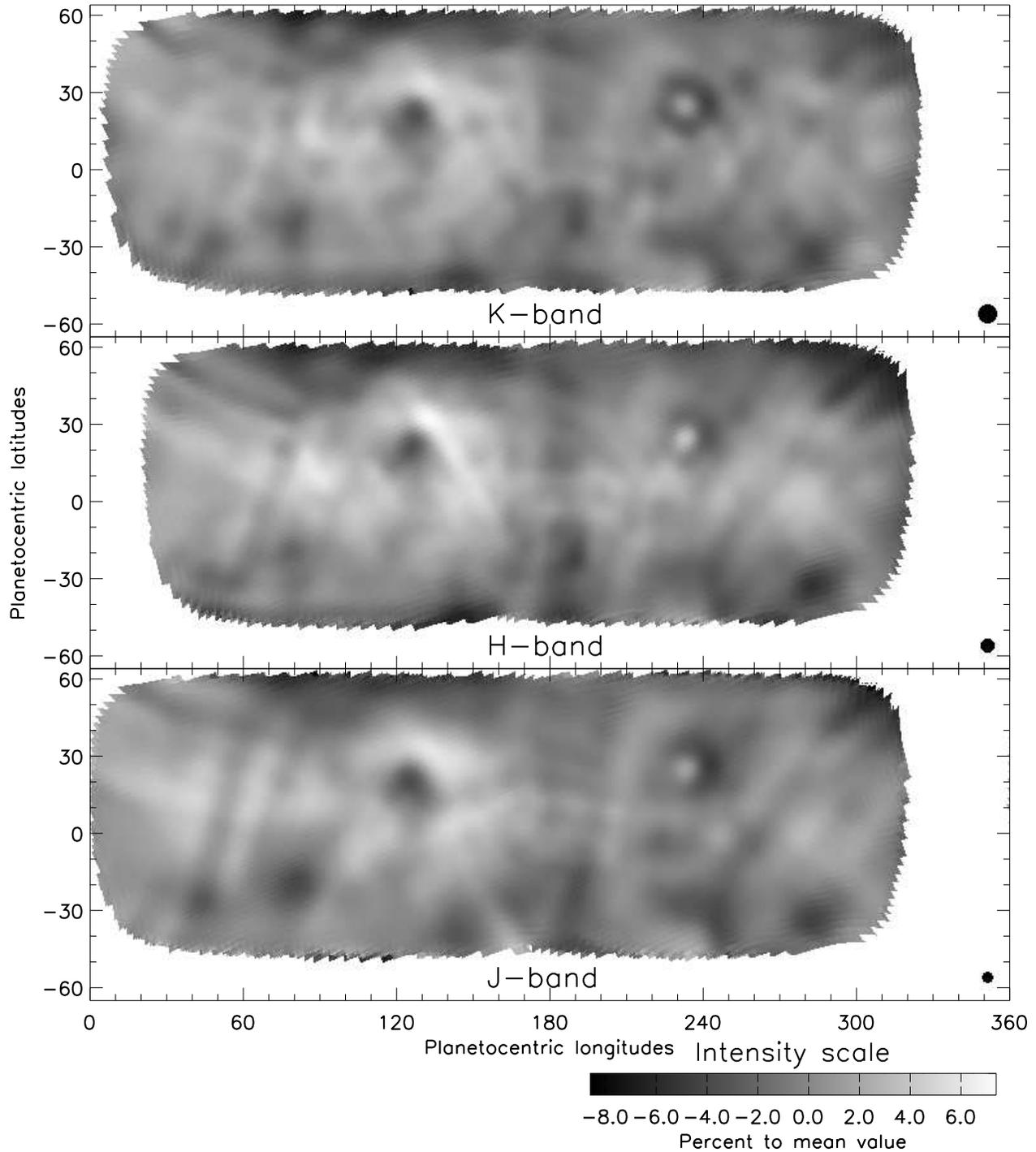}
    \caption{J-, H- and K- band maps of Ceres covering $\sim$80\% of
    Ceres' surface. The areas in white are \textsl{terra incognita}
    due to the limited ROI and rotational phases not imaged from Keck
    (see text). The theoretical resolution elements $\Theta$ at J/H/K bands are
    shown at the bottom-right corners. Albedo variations are within
    $\pm$6\% around the mean surface value for each map. We estimate
    the errors to be 2.5\% maximum (see
    \textbf{Fig. \ref{fig-error_maps}}). The color scale is common for
    the three filters. Several round shaped features are visible
    including a dark feature with a bright center spot at
    $\sim$(234\degr, +23\degr) (named ``A'') and a dark region at
    $\sim$(125\degr, +20\degr) named ``Piazzi'' by
    \citet{2002-AJ-123-Parker} (here ``B''). Two other dark features
    are visible in the Southern hemisphere at $\sim$(80\degr,
    -20\degr) and $\sim$(285\degr, -35\degr) as well as other smaller
    features elsewhere (see \textbf{Fig. \ref{fig-spectra}}). Any
    feature or albedo distribution present in the three maps has a
    very low probablity to be an artifact, with the exception of the
    diagonal stripes seen across the surface in H- and J-band (see
    text and \textbf{Fig. \ref{fig-error_maps}}). One hemisphere
    (0\degr~to 180\degr) appears to be sensibly brighter (1\%) than
    the other at these wavelengths. A dark region running North/South
    is located at the boundary between the two hemispheres and is
    present in all maps.}
    \label{fig-jhk-maps}
\end{figure*}

%%%%%%------ FIGURE --- Begin --- J-/H-/K-band Maps ------%%%%%%
%
%
%
%
    \indent To better represent the distribution and spatial extent of the features observed on the surface of Ceres (albedo, geological marks), we projected our high-angular resolution images into maps. The following sections are used to describe in details the various steps involved in the process of map projection, and the subsequent analysis of Ceres surface properties. 

%
%
%
%%%%%%------ Cylindrical Equidistant Projection Choice ------%%%%%%
    \paragraph*{Geometry:}~Any projection of an ellipsoidal shape onto a plane introduces deformations \citep{SurfaceMapping}. We attempted to minimize these effects by choosing the cylindrical equidistant projection, which maps the surface of the asteroid onto a cylinder tangent to its equator and conserves the distances along the meridians. As a result, this projection minimizes the deformations of Ceres' equatorial area, which is seen at highest resolution thanks to its small obliquity. Higher planetocentric latitudes suffer stronger deformations after projection, but the impact is mitigated since these regions correspond to areas imaged tangentially, at a lower equivalent spatial resolution. 
%
%
%
%
%%%%%%------ Definition of a ROI ------%%%%%%
    \paragraph*{Region of interest:}~We produced albedo maps of Ceres using the pixels located within an ellipsoidal Region Of Interest (ROI) centered on the image, and whose semi-axes were equal to 80\% of the corresponding projected semi-axes on Ceres. The ROI corresponds to 64\% of the projected surface for each image (see \textbf{Fig. \ref{fig-3DvsIM}}). Ignoring the pixels near the edge of Ceres was based on several considerations:
    \begin{itemize}
      \item[1-]Although all images used to produce the maps were carefully cleaned before deconvolution (including from correlated noise), some of them still presented artifacts after restoration of their optimal resolution, particularly near discontinuities such as the limb and terminator.
      \item[2-]It appeared difficult in a few cases to restore optimally both the contour of the object and the surface details, even if \textsc{Mistral} is optimized to minimize the ``ringing" effect introduced by the deconvolution of sharp edge objects.
      \item[3-]Finally, the resolution per pixel being highest at the
        center of the disk of Ceres, the
        use of the pixels located
            near the edge would strongly degrade the optimal resolution of our final product.
    \end{itemize}
%
%
%
%
%%%%%%------ Luminuous gradient correction ------%%%%%%
    \paragraph*{Phase angle correction:}~As seen on \textbf{Fig. \ref{fig-views-of-ceres}}, a phase angle of just a few degrees at near-infrared wavelengths produces a strong gradient of luminosity across the disk of Ceres. This effect had to be corrected prior to combine our individual images onto single maps. Several diffusion laws were investigated to model it, such as adopting a simple linear gradient, or using more complete models such as provided by the Lambert, Lommel-Seelinger, Minnaert and Hakpe laws (including single-scattering and multiple-scattering effects) \citep[see][]{hapke-theory}. The linear gradient was adopted since it provided the best fit to Ceres surface (residuals produced were nearly twice smaller than in the case of the Hapke model).
%
%
%
%
%%%%%%------ From (x,y) frame to (lambda,phi) one ------%%%%%%
    \paragraph*{Projection:}~The main difficulty to project an image of Ceres into its planetocentric referential (defined by its planetocentric longitude ($\lambda$) and latitude ($\varphi$) as recommended by the IAU \citep{2005-CeMDA-91-Seidelmann}), resides in the accurate determination of its geometrical center. As the distribution of Ceres gravity field is unknown, we made the assumption that its center of mass, which is the center of the planetocentric referential, coincides with its geometrical center. The conversion into planetocentric coordinates was based on our determination of Ceres dimensions (see \ref{subsec-dimensions}) and the orientation of its spin axis (SVC, see \ref{subsec-pole-coordinates}), which is defined by the north pole angle ($p_n$) and by the Sub-Earth Point coordinates (SEP$_\lambda$, SEP$_\varphi$). We used our value for Ceres rotation period (see \ref{subsec-rotation}) and the Eproc ephemeris generator to obtain the SEP$_\lambda$, SEP$_\varphi$ and $p_n$ at the time of each observation. We then projected the images onto half-degree gridded maps to sample all areas of Ceres with sufficient resolution, from the equatorial region, up to the edge of the ROI. This fine grid map helped recover the smallest scale information from our set of 360 images.
%
%
%
%
%%%%%%------ Worldwide map from individual ones ------%%%%%%
    \paragraph*{Combination of images into maps:}~No absolute
    photometric calibration was obtained for our data. We therefore
    used a near-infrared disk-integrated spectrum of Ceres [R.Binzel,
      personal communication] normalized to unity at 0.5 $\mu$m to
    calculate the equivalent disk-integrated photometric value for
    each band, and normalize our albedo maps with respect to each
    other. Prior to combine the different views into maps for each
    wavebands, we corrected the intensity variations of the individual
    projections caused by the differential atmospheric absorption
    across our wavelength range, as well as the quality of the AO
    correction. We first projected each image onto the planetocentric
    referential of Ceres. Then we adjusted the brightness level of
    spatially adjacent maps by measuring the flux ratio over their
    overlapping area and applying the corresponding re-normalization
    coefficient. After correction, the maps could be combined to
    produce the albedo maps presented in
    \textbf{Fig. \ref{fig-jhk-maps}} for each waveband. Each pixel of
    the final maps (covering a quarter of a square degree) was
    obtained by combining the corresponding pixels from the individual
    projections using a gaussian-weighted average function (we chose
    a gaussian with a standard deviation of 5 pixels). The largest
    weight (weight value = 1) was attributed to the pixel
    providing the best spatial resolution, while the pixels with a
    lower resolution were assigned a lower weight. \textbf{Table
      \ref{tab-pixmap}} represents the average and maximum number of
    images used to produce a single pixel of the final maps for each
    waveband, as well as the percentage of surface coverage of Ceres
    ($\sim$80\%). In order to facilitate comparison between our
    near-infrared data and the UV/Visible (223, 335 and
    535 nm) HST maps, we chose to adopt the same
    reference meridian than published by \citep{2006-Icarus-182-Li}.
%
%
%
%
%%%%%%------ TABLE --- Begin --- Maps per Pixel ------%%%%%%
\begin{table}
  \centering
  \begin{tabular}{c c c c}
    \hline\hline
    Filter &   Average Images  &   Maximum Images  &  Coverage \\
           &  (\# per pixel)   & (\# per pixel)   & (\% total surface) \\
    \hline
    J   &   16  &   63  &  81.8  \\
    H   &   14  &   54  &  74.2  \\
    K   &   19  &   72  &  80.0  \\
    \hline
\end{tabular}
  \caption{Average and maximum number of images used to produce a single pixel of the final J/H/K bands maps, and their corresponding surface coverage.}
  \label{tab-pixmap}
\end{table}
%%%%%%------ TABLE ---  End  --- Maps per Pixel ------%%%%%%
%
%
%
%
%
%%%%%%------ FIGURE --- Begin --- error maps------%%%%%%
\begin{figure}
  \resizebox{\hsize}{!}{\includegraphics{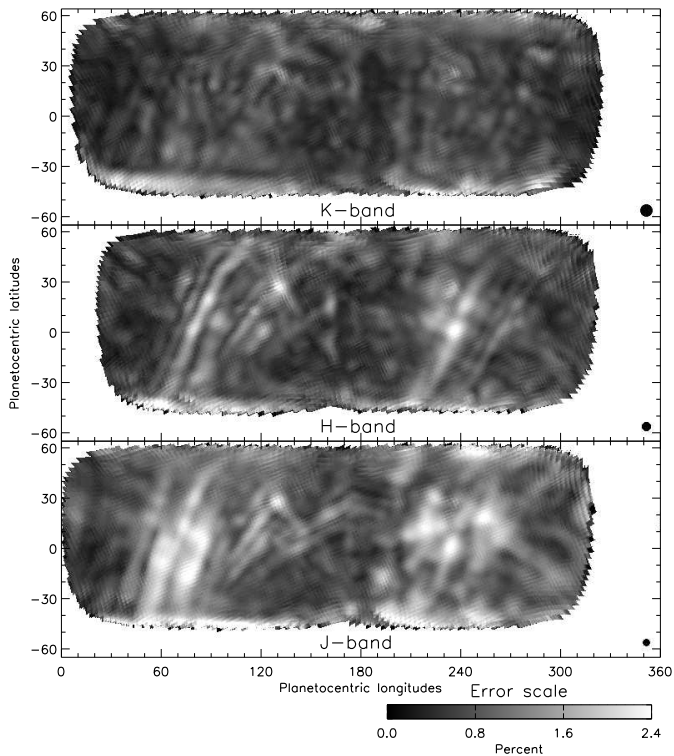}}
  \caption{1-$\sigma$ deviation map expressed in percent with respect
    to the map mean value for each filter. The error is greater in
    J-band than in K-band as it could be expected from the lower
    strehl ratio at smaller wavelengths. The error also grows with
    latitude, which is due to the nearly equatorial view of Ceres as
    seen from an Earth observer.}
  \label{fig-error_maps}
\end{figure}
%%%%%%------ FIGURE --- End --- error maps ------%%%%%%
%
%
%
%
%
%%%%%%------ Maps description and informations ------%%%%%%
    \paragraph*{Maps description:}~The J-, H- and K-band maps shown in
    \textbf{Fig.~\ref{fig-jhk-maps}}, and covering $\sim$80\% of
    Ceres' surface (see \textbf{Table \ref{tab-pixmap}}), are the
    result of combining 126, 99 and 135 individual projections
    respectively. 
    We also derived error albedo
    maps (\textbf{Fig. \ref{fig-error_maps}}) by measuring, for each
    pixel, the intensity dispersion across the individual views.
    The theoretical size of the resolution element for
    J-, H- and K-band is 36.8 km, 47.4 km and 62.9 km
     respectively (corresponding to 4.4\degr, 5.6\degr~and 7.5\degr~at the
    equator). The major features sustain diameters of $\sim$180 km (A
    and B) but smaller features can be seen in all three maps down to
    $\sim$50 km scale. Although the theoretical resolution is highest
    in J-band, \textbf{Fig. \ref{fig-jhk-maps}} shows that the final
    resolution is nearly equivalent across our three bands
    (\textsl{i.e.} $\sim$60 km at equator). The degradation of the H-
    and J-band resolution is due to a more variable PSF at these
    wavelengths, which is supported by the larger photometric error
    (\textbf{Fig. \ref{fig-error_maps}}) derived for these bands. The
    amplitude of the albedo variation is within $\pm$6\% around the
    mean surface value for each band. The
    error maps show that the albedo maps
    (\textbf{Fig. \ref{fig-jhk-maps}}) display an increasing error
    with decreasing wavelength: the 1-$\sigma$ uncertainty is smaller
    than 1\% in the equatorial area in K band while it is estimated to
    be $\sim$2\% in J-band. The shape of the error distribution around
    60\degr~longitude reveal that the linear oblique features seen in
    J- and H-band are noise and should be ignored from our
    analysis. The K-band map displays the lowest noise level, due to
    the highest Strehl ratio delivered by the AO system at these
    wavelengths and should be considered as the most accurate of the
    three.\\
    \indent Some of the main albedo features in our near-infared maps
    can also be seen in the UV/Visible HST maps published by
    \citet{2006-Icarus-182-Li}, like the
    large bright area around (125\degr, +20\degr) and the dark spot
    at (130\degr, -24\degr).
    The fact that these features are
    visible in all wavelengths suggests that they are
    geological features like basins or impact craters...
    But the UV/Visible and near-infrared maps show also
    discrepancies: the bright feature at (115\degr, -30\degr) visible
    in the UV/Visible maps is not present in our maps, nor the dark feature
    located around (45\degr, +10\degr).
    Whereas all the discrepancies cannot be linked to surface
    properties (the dark annulus of the ``A'' feature (234\degr,
    +23\degr) is not visible
    in the UV/Visible map due to its size which is comparable to the resolution element provided by
    the HST), such variations in their spectral behavior suggest compositional
    differences between these regions. The next section
    will provide a discussion on possible composition and geological
    origin.\\
  \subsection{3-D model}
%
%%%%%%------ FIGURE --- Begin --- 3D VERSUS IMAGE------%%%%%%
\begin{figure}
%%%--- Graphics from the EPS and Caption
  \resizebox{\hsize}{!}{\includegraphics{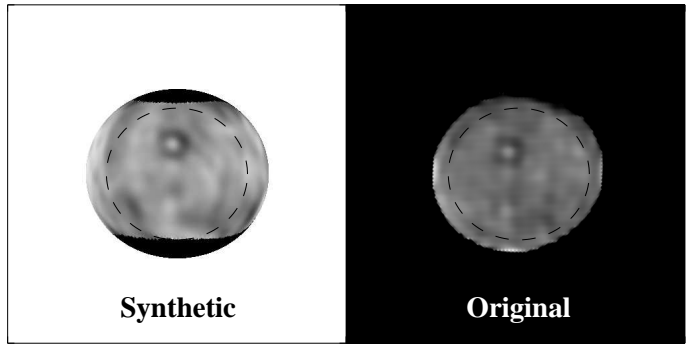}}
%%%--- Nom des filtres

  \setlength{\unitlength}{\hsize}
  \begin{pspicture}(0.1,0.01)
   %%-Refereee version
%    \setlength{\unitlength}{1.55cm}
    \put(1.5,0.7){\large{\textbf{Synthetic}}}
    \put(6.1,0.7){\large{\textcolor{white}{\textbf{Original}}}}
  \end{pspicture}

  \caption{Comparison between a synthetic view of Ceres (left) and 
    a single deconvolved image (right). The overplotted dotted ellipse
    corresponds to the ROI limit (80\% of projected axis). The synthetic view appears ``smoother''
    than the original view because it is the result of the weighted
    average of several deconvolved images. The main albedo features
    are seen in both the synthetic and the original views, while other
    features did not resist the
    weighted average of several images, which acted as a filter to
    remove artifacts present in a single frame.}
  \label{fig-3DvsIM}
\end{figure}
%%%%%%------ FIGURE --- End --- 3D VERSUS IMAGE ------%%%%%%
%
%
%
\footnotetext[1]{
\href{http://www.imcce.fr/page.php?nav=en/ephemerides/formulaire/form\_ephephys.php}
    {http://www.imcce.fr/page.php?nav=en/ephemerides/formulaire/\\
     form\_ephephys.php}}
    \indent We re-projected our multicolor maps of Ceres onto its 3D
    shape model. Such model is used to check the distribution of
    albedo features present in our final maps against the features
    seen in our original images after deconvolution (see
    \textbf{Fig. \ref{fig-3DvsIM}}). This model can also be used to
    predict\,\footnotemark[1] what Ceres surface would look like at
    any epochs, which will certainly be useful when preparing
    any future ground-based, or space based observations of Ceres.\\
  \subsection{Three-band spectra}
\begin{figure}
 \centering
   \resizebox{.9\hsize}{!}{\includegraphics{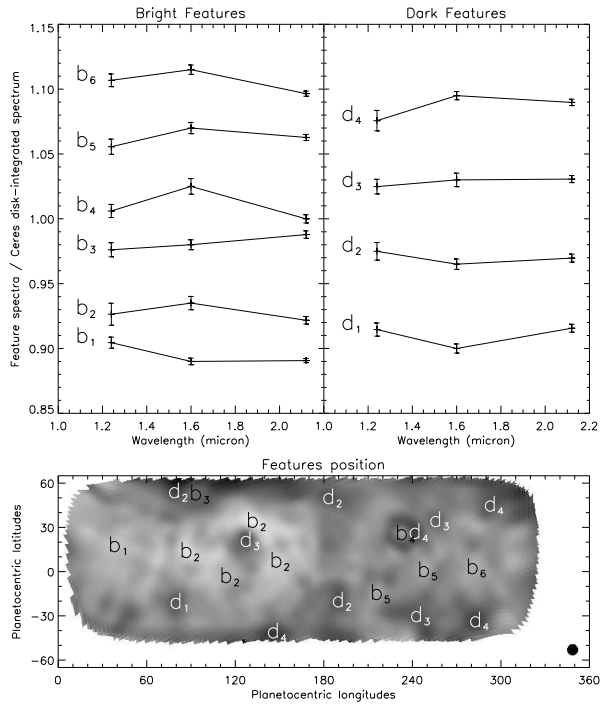}}
  \caption{Three-band spectra of selected surface features normalized
    to Ceres disk-integrated spectrum. The left
    (respectively right) panel shows the
    photometric points obtained for the bright
    (respectively dark) features. All
    spectra are shifted vertically by 0.045
    (respectively 0.065)
    to improve clarity. The letters
    positioned left of each spectrum are used to locate the
    corresponding features on the map. The K-band resolution element
    is shown at the bottom-right corner.}

  \label{fig-spectra}
\end{figure}

    \indent We investigate the nature of the major features seen in
    our maps by reporting their relative brightness variation at J-,
    H- and K-bands. We selected 10 type of area (6 bright and 4 dark,
    noted b$_i$ and d$_i$ respectively)
    and represented in \textbf{Fig. \ref{fig-spectra}} their
    photometric measurements normalized to Ceres integrated
    spectrum. The photometric value obtained for each band is the
    result of averaging the photometric measurements over an area
    equivalent to a resolution element. We estimate the error on the
    relative photometry as $\sqrt{\sum \sigma_i^2}$, where $\sigma_i$
    is the standard deviation for each pixel as read in the error map
    (\textbf{Fig. \ref{fig-error_maps}}). The results are shown in
    \textbf{Fig. \ref{fig-spectra}} with their 3-$\sigma$ error
    bars.\\
    \indent Whereas spectral variation with the rotational phase of
    Ceres has never been reported, our data show differences of
    spectral behavior across the surface. The analysis of
    \textbf{Fig. \ref{fig-spectra}} reveals a clear trend of the
    bright features (left) to display a higher H-band albedo (with
    respect to J- and K-band) than the rest of Ceres surface. On
    the contrary, the dark features (right) do not display a similar
    trend. This may indicate a common origin for the bright features,
    whereas the dark regions may be the result of various surface
    processes, or represent different level of surface aging. \\
\footnotetext[2]{
    \href{http://speclib.jpl.nasa.gov}{http://speclib.jpl.nasa.gov}}
\footnotetext[3]{
    \href{http://www.planetary.brown.edu/relab/}{http://www.planetary.brown.edu/relab/}\\
    \textcolor{white}{\hspace{2cm}space for the url}}

\begin{figure}
 \centering
   \resizebox{.7\hsize}{!}{\includegraphics{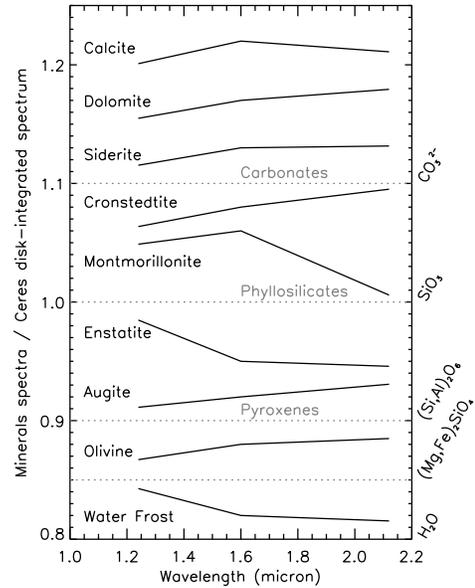}}\\

   \caption{Three-band spectra of selected
     carbonates [CO$_3^{2-}$],
     phyllosilicates [SiO$_3$],
     pyroxenes [(Si,Al)$_2$O$_6$],
     olivine [(Mg,Fe)$_2$SiO$_4$],
     and water ice frost [H$_2$O]
     normalized to Ceres disk-integrated spectrum. Error bars are negligible.}
   \label{fig-spectra-minerals}
 \end{figure}

    \indent We then compare the 3-band spectra to those of various
    laboratory compounds. We used the ASTER\,\footnotemark[2]
    and RELAB\,\footnotemark[3]
    spectral libraries, to
    obtain the equivalent near-infrared spectra of various compounds
    predicted to be present on Ceres. We report in
    \textbf{Fig. \ref{fig-spectra-minerals}} their broad band
    photometric values, normalized to Ceres disk-integrated
    spectrum, each mineral being mixed with a dark neutral compound, using the
    mixing ratios given in \textbf{Table \ref{tab-mineral}}.
    Due to the low spectral resolution provided by our broad band imagery, 
    it is not possible to identify unequivocally the compounds present on 
    the surface of Ceres. We thus based our study to the comparison between the 
    possible surface compounds proposed by \citet{2006-Icarus-185-Rivkin},
    who showed that carbonates like Siderite, Dolomite or Calcite
    mixed with phyllosilicates provide an
    excellent fit to the 3 $\mu$m region of Ceres. Carbonates could
    also explain the mid-infrared spectral emission features detected
    by \citet{1998-AJ-115-Cohen}. 
    We also included an orthopyroxene (Enstatite), a clinopyroxene
    (Augite) and Olivine, the most abundant elements in the Solar
    System. A water ice frost spectrum is also reported.\\
%
%
%
%%%%%%------ FIGURE --- Begin --- 3-point spectra ------%%%%%%
%%%%%%------ FIGURE --- End --- 3-point spectra ------%%%%%%
%
%
%%%--- First: Carbonates + Phyllosilicates do match well bright features
    \indent Although the compositional information returned at such a
    low spectral resolution should be considered with caution,
    we can use the comparison between the
    spectral behavior measured on the surface and that of the
    laboratory samples to constrain and discuss the possible surface composition
    of Ceres. There are similarities between the spectral behavior of the
    bright features seen in \textbf{Fig. \ref{fig-spectra}} (left) and
    that of phyllosilicates and carbonates in
    \textbf{Fig. \ref{fig-spectra-minerals}}. The Calcite and
    Montmorillonite ``spectra" display the same shape as the majority
    of the bright features. For instance, the 
    Montmorillonite reproduces quite well the behavior of the b$_4$
    feature as well as the bright region b$_2$ surrounding
    d$_3$. 
    On the other hand, the spectral behavior of
    the Siderite (a iron-rich Calcite), 
    Cronstedtite (an iron-bearing phyllosilicate), or
    Augite (a clinopyroxene)
    only match the
    spectrum of a small percentage of Ceres surface, and for small mixing
    ratios show that all of them might be minor
    compounds of the regions discussed in this section. Although,
    igneous rocks like pyroxenes and olivine are not expected to be
    present on the surface of primitive asteroids, these compounds are
    likely minor components of the regions discussed in this section
    if present at all.\\
%
%
%
%
%
%
%
%%%%%%------ FIGURE --- Begin --- Mineraux spectra ------%%%%%%
%%%%%%------ FIGURE --- End --- Mineraux spectra ------%%%%%%
%
%
%
%%%%%%------ TABLE --- Begin --- SpecLIB ------%%%%%%
\begin{table}
  \centering
  \begin{tabular}{cccc}
    \hline\hline
    Component &   Grain Size & Mixing ratio &  Sample\\
              &   ($\mu$m)   &  (\%)        & No.\\
    \hline
    Calcite\,$^{a}$         &   0-45   & 100 & C-3A       \\ %Ward's               \\%(CaCO$_3$)
    Dolomite\,$^{a}$        &   0-45   & \textcolor{white}{0}50 & C-5A  \\ %Ward's  %(CaMg(CO$_3$)$_2$
    Siderite\,$^{a}$        &   0-45   & \textcolor{white}{00}4 & C-9A  \\ %JPL %(Fe$^2$+CO$_3$)
    Crondstedtite\,$^{r}$   &   0-45   & \textcolor{white}{00}5 & CR-EAC-021 \\ %Edward A. Cloutis    \\
    Montmorillonite\,$^{a}$ &   0-45   & \textcolor{white}{0}50 & PS-2D  \\ %Clay
                                                             %Mineral Society
                                                             %\\%(Na,Ca)$_0.33$(Al,Mg)$_2$Si$_4O_10$(OH)$_2$.H$_2$O
    Enstatite\,$^{a}$       &   0-45   & \textcolor{white}{0}30 & IN-10B     \\ %Mg$_2$(SiO$_3$)$_2$
    Augite\,$^{a}$          &   0-45   & \textcolor{white}{00}2 & IN-15A     \\ %(Ca,Na)(Mg,Fe,Al,Ti)(Si,Al)$_2$O$_6$
    Olivine\,$^{r}$         &   0-50   & \textcolor{white}{00}5 & DD-MDD-013 \\ %Melinda D. Dyar
    Water Frost\,$^{a}$     & \,$\sim$10 & \textcolor{white}{00}2 & FROST.SNW \\ %John Hopkins Univ.   \\%H$_2$O
    \hline
  \end{tabular}
  \caption{Grain size and mixing ratio for the selected compounds. The
    last column indicates the sample reference in the spectral
    libraries ($a$: ASTER and $r$: RELAB).}
  \label{tab-mineral}
\end{table}
%%%%%%------ TABLE ---  End  --- SpecLIB ------%%%%%%
%
%%%--- Second: Dark Features are kind of a problem, due to low-H + many shapes
    \indent The interpretation of the  dark features is more
    complex. Indeed, the lower H-band value that characterizes half of
    the dark features (d$_1$, d$_2$)
    does not match the behavior seen in phyllosilicates and
    carbonates, which are the major compounds predicted to be present
    over Ceres (see \ref{sec-review-water}). Such a drop in H-band
    matches better the behavior of 
    Enstatite, but its high density of $\sim$3\,200
    kg.m$^{-3}$ and the mid-infrared
    mismatch between Ceres and Enstatite spectra makes its presence unprobable.
    Another possible matching compound is water ice
    (\textbf{Fig. \ref{fig-spectra-minerals}}). Water ice is
    not expected to be found in a stable form over
    Ceres, but ``dirty'' ice, \textsl{i.e.} ice or frost
    mixed with other minerals, could be present and reproduce the more elevated
    K-band point seen in the ``spectra'' of these features. Another point to consider is
    the relatively high planetocentric latitude of these dark
    features. None of them is found in the equatorial region, most
    being located above 30-40\degr~latitude. If dirty ice exists on
    the surface or sub-surface layers of Ceres
    \citep{2005-JGR-110-McCord, 2005-MNRAS-358-Mousis}, it would be
    expected to be more stable at higher latitudes, where surface
    temperature is lower. Nevertheless, we cannot ascertain from these
    data alone that ice is present on Ceres.
    Higher spectral resolution, coupled with the high-contrast and spatial resolution
    provided by adaptive optics, is required to investigate in detail
    the composition of the main features seen on its surface. The
    limits of such a broad-band analysis in constraining the
    composition of the main features is illustrated by the bright feature b$_1$, which
    appears to match quite well the behavior expected for water frost,
    but it is located at low latitude and differ in reflectivity from
    the dark features discussed above. \\
%
%%%--- A and B feature --- Highly speculative
    \indent The two main observed features A (b$_4$, d$_4$) and B
    (d$_3$ and the surrounding region b$_2$), were both
    referenced as \#5 and \#2 by \citet{2006-Icarus-182-Li}. These two
    features are remarkable because whereas they sustain large
    physical dimensions ($\diameter_\textrm{A} \sim 180$ km and
    $\diameter_\textrm{B} \sim 350$ km), their spectral behavior appears to
    be homogeneous over such large area, which might point to a same
    composition and/or resurfacing history. The external annulus of
    the B feature (b$_2$) shows the same uniformity. This
    annulus shape is reminiscent of a large cratering event. The
    bright central regions of feature A resembles the central peak
    seen in craters originating from high energetic impact, which
    would then be subject to different aging processes than the
    lower altitude neighboring areas. At smaller physical  
    scale, similar differences of albedo have been reported on asteroid  
    25143 Itokawa between the central part and surrounding areas of small  
    craters, this time via grain sorting \citep{2007-Science-316-Miyamoto}.\\
    \indent In summary, our AO study permits to map the albedo
    variations over the surface of Ceres down to a 40-60 km scale and
    investigate whether these variations correspond to changes in the
    composition. Also, the high-spatial resolution capabilities
    returned by AO make possible to search for areas of distinctive
    signature (e.g. icy rich spots), which would remain undetected
    otherwise in disk-averaged studies of Ceres. A dusty regolith a
    few centimeters thick \citep{1988-AJ-95-Webster}, created by the
    impact of micrometeorites and possibly larger bodies, has been
    proposed \citep{1999-AJ-117-Witteborn,2005-Icarus-173-Lim} as a
    possible explanation for the shallow
    spectral signatures in the visible and
    near-infrared spectrum of Ceres and its small
    albedo constrast. But some small areas visible in our
    high-angular resolution images of Ceres could correspond to places
    on the surface where the regolith material has been cleared by
    ``recent'' endogenic or exogenic activity to expose more pristine
    material from the sub-surface layers.

  \section{Conclusions}
  \indent We present the results of high-angular resolution
  near-infrared ($\sim$[1.1,2.3] $\mu$m) imaging observations conducted with the adaptive
  optics system at the Keck II telescope in Hawaii. Imaging Ceres over
  nearly one rotation period allows us to derive its main physical
  characteristics: rotation period ($P_s = 9.074\,10_{-0.000\,14}^{+0.000\,10}$ h),
  spin vector orientation (EQJ2000.0 $\alpha_0 = 288\degr \pm 5$\degr, 
    $\delta_0 = +66\degr \pm 5$\degr),
  size ( $a = b = 479.7 \pm 2.3$ km, $c = 444.4 \pm 2.1$ km ) and
  density ($\rho=2\,206 \pm 43$ kg.m$^{-3}$)
  and produce the first near-infrared maps of its surface, revealing
  a wealth of surface features (also seen in the UV/Visible wavelengths \citep{2006-Icarus-182-Li}.
  We discuss the nature and composition of the main albedo
  marks present on its surface. We do not detect topographic features
  over 18 km high, which agrees with predictions for a body of
  this size whose internal composition could still contain a large
  amount of water ice. By comparison with laboratory samples, we found
  Ceres surface composition to be consistent with 
  phyllosilicates such as Montmorillonite, and carbonates such as
  Dolomite or Calcite. Some dark regions above
  30-40\degr~latitude display the characteristics expected from
  water-frost mixed with minerals. Unfortunately, broad band imaging
  does not provide sufficient spectral resolution to investigate
  further the composition of distinct geological features present on
  Ceres, and future spectroscopic observations combined with high-angular
  resolution are required to better constrain its mineralogy.
 \section*{Acknowledgments}
  \indent This work was supported by
  a grant from the NASA Planetary Astronomy program (RTOP
  344\_32\_55\_05, PI:
  C. Dumas). We would like to thank Rick Binzel for providing us Ceres near-infrared
  spectrum, Mark Sykes for providing Ceres rotational data
  ahead of publication, and the Keck II Observatory team for their support. These
  observations were performed on NASA Keck Time. This research uses
  spectra acquired by Edward A. Cloutis and by Melinda D. Dyar
  with the NASA RELAB facility at Brown University.

%\footnotetext[1]{http://www.imcce.fr/page.php?nav=en/ephemerides/formulaire/form\_ephephys.php}

%%%--- Bibliography ---%%%
\bibliographystyle{aa}      %-Astronomy And Astrophysics Style 
\bibliography{biblio}

\end{document}